\documentclass[a4paper,12pt]{article}

\usepackage{amsmath}
\def\M{ \overline{|\mathcal{M}|^2} }
\def\gluino{\mathaccent"7E g}  \def\mgluino{m_{\gluino}}
\def\squark{\mathaccent"7E q}  \def\MiLR{M_{i_{L,R}}}
\def\gsf{g_s^4}                \def\nc{\ensuremath{N_C}}
\def\cf{\ensuremath{C_F}}      \def\ca{\ensuremath{C_A}}
\def\np{\ifmmode{\rm n.p.}\else{nonplanar}\fi} \def\lr{\leftrightarrow}
\def\qcd{\textsc{Qcd}}         \def\susy{supersymmetric}
\def\MC{Monte Carlo}           \def\herwig{\texttt{HERWIG}}
\newcommand{\cavendish}[1]{{\it preprint} Cavendish--HEP--#1}
\newcommand{\xxx}[3]{ {\bf #1} (19#2) #3}
\newcommand{\plb}[3]{{\it Phys.\ Lett.}\ \xxx{B#1}{#2}{#3} }
\newcommand{\prd}[3]{{\it Phys.\ Rev.}\  \xxx{D#1}{#2}{#3} }
\newcommand{\npb}[3]{{\it Nucl.\ Phys.}\ \xxx{B#1}{#2}{#3} }
\newcommand{\cpc}[3]{{\it Comp.\ Phys.\ Comm.}\ \xxx{#1}{#2}{#3} }

\begin{document}
\thispagestyle{empty}
\setcounter{page}{0}

\begin{flushright}
{\large Cavendish--HEP--98/05}\\ {\rm June 1998}\\ \end{flushright}
\vspace*{\fill}\begin{center}
{\LARGE \bf Colour Connection Structure of \\[0.5cm]
            (Supersymmetric) QCD \\[0.5cm]
            ($\bf 2\to2$) Processes \\[2.cm] }
{\Large Kosuke Odagiri} \\[3mm]
{\it Cavendish Laboratory, Madingley Road, Cambridge CB3 0HE, UK} \\[5mm]
\end{center}

\vspace*{\fill}\begin{abstract}{\small\noindent
The colour connection structure of \qcd\ ($2\to2$) processes is discussed,
with emphasis on its application to the \susy\ 2 parton $\to$ 2 sparton
processes, which are currently being implemented in the \herwig\ \MC\
event generator. The procedure described by Marchesini and Webber is found
to be inadequate, and a new method is proposed. However, this alteration
is unlikely to significantly affect the theoretical predictions for soft
gluon radiation. A complete list of \susy\ \qcd\ $2\to2$ matrix elements
and their colour decompositions is presented.  }\end{abstract}
\centerline{{\it PACS}: 12.38.-t; 12.38.Bx; 12.60.Jv; 14.80.Ly}
\vspace*{\fill}\newpage\setcounter{page}{1}

\section{Introduction}\label{sec_intro}

The simulation of soft gluon radiation in hard (\susy) \qcd\
processes \cite{MW,EMW,WLHC,MOSW} requires that the corresponding matrix
elements be rearranged according to the colour connections (defined by the
colour flows) in the process \cite{MW}. In brief, this is because the
colour connections in the parent process determine the cones in which the
soft gluons radiate and hence the hadronisation occurs.

This rearrangement of terms is automatic if there is a unique colour flow
associated with the process, as is the case for the \qcd\ $qq'\to qq'$
scattering whose colour flow is shown in figure \ref{figA}, but for
more complex processes the procedure involves some ambiguity. The purpose
of this paper is to discuss and analyse this ambiguity, to propose a
consistent and practicable method, and to illustrate the application of
this new method in \susy\ \qcd\ 2 parton $\to$ 2 sparton processes which
are being implemented in the \herwig\ \MC\ event generator
\cite{MOSW,H59,H61}.

\subsection{Colour flows in QCD}\label{subsec_col}

The matrix elements for processes with more than one colour flow consist
of the `planar' terms and the `\np' terms. The planar terms are
those with single colour flows and the \np\ terms are those with no
single colour flow. The \np\ terms are always suppressed by some
inverse powers of \nc. 

The colour flows for the four distinct $2\to2$ \qcd\ processes are shown
in figures \ref{figA}--\ref{figD}. For concreteness, let us consider the
\qcd\ process $q\bar{q}\to gg$, for which the leading-order spin- and
colour-averaged matrix element squared is given by \cite{CKR}:
\begin{equation} \gsf\cf\frac{t^2+u^2}{s^2}\left[
\left(\frac{u}{t}\right)_t+\left(\frac{t}{u}\right)_u+
\left(-\frac{1}{\nc^2}\cdot\frac{s^2}{ut}\right)_\np
\right]. \end{equation}
The first term in braces is a planar term corresponding to the $t$-channel
colour flow, and the second term corresponds to the $u$-channel colour
flow, as depicted in figure \ref{figC}. The third term, suppressed by the
factor $(1/\nc^2)$, is a \np\ contribution corresponding to a mixed colour
flow, which needs more care in its treatment when we consider the
rearrangement according to the colour connection.

Note that apart from the overall gauge invariance and positive
definiteness of the matrix element squared, each of the above three terms
is also gauge invariant and, in the case of the planar terms, positive
definite. The gauge invariance follows from the fact that colour is
formally an observable at Born approximation. As for
the positive definiteness of the planar terms, this is obvious since the
modulus squared of any part of the full matrix element must also be
positive. This gauge invariance allows us to uniquely identify the planar
terms and the \np\ part in each process.

\subsection{Radiation from colour connected partons}\label{subsec_rad}

Let us recall the results of Ellis, Marchesini and Webber \cite{MW, EMW}
concerning the coherence of soft gluon radiations \cite{TEV1, TEV2} in
hard \qcd\ processes.

For each colour flow (each planar term) there is a cone around each
incoming and outgoing parton direction bounded by the angle between the
parton and the parton which is colour connected to it\footnote{This
statement has not been explicitly verified in the \susy\ case involving
either long-lived spartons in the final state (the light gluino, whose
existence is still controversial --- see \cite{FGLU}.) or short-lived ones
in the intermediate state.}. In the case of gluons (and gluinos) there are
two such cones, one for the colour and one for the anticolour. The cones
define the bounds for the soft gluons to be radiated (in the angular
ordering approximation) and hence the hadronisation\footnote{When the
(s)partons are massive, the analysis needs some modification \cite{HEAVY}.
For the heavy quark case, there will be a `screening' of the collinear
direction \cite{WLHC}. For the \susy\ case, the analysis has not yet been
carried out.} to occur, to leading order in \nc. 

As for the \np\ part, this can be distributed in any way between the
colour flows. Let us consider this in detail.

Returning to the example of $q_1\bar{q}_2\to g_3g_4$ introduced earlier,
the radiation pattern, in the notation of \cite{MW}, is:
\begin{equation} \gsf\cf\frac{t^2+u^2}{s^2}\left[
\left(\frac{u}{t}\right)_tW_t+\left(\frac{t}{u}\right)_uW_u+
\left(-\frac{1}{\nc^2}\cdot\frac{s^2}{ut}\right)_\np W_\np
\right], \end{equation}
with:
\begin{eqnarray}
W_t   &=& \ca\left[W_{34}+W_{13}+W_{24}-W_{12}\right]
        +2\cf      W_{12} \\
W_u   &=& \ca\left[W_{34}+W_{14}+W_{23}-W_{12}\right]
        +2\cf      W_{12} \\
W_\np &=& \ca\left[W_{13}+W_{24}+W_{14}+W_{23}-2W_{12}\right]
        +2\cf      W_{12}. \end{eqnarray}
\cf\ and \ca\ are as usual the \qcd\ colour factors $(\nc^2-1)/2\nc$ and
(\nc) respectively. Terms proportional to \ca\ correspond to the radiation
from the gluon legs, whereas the terms proportional to 2\cf\ correspond to
the radiation from the quark legs. $W_{ij}$ are the radiation functions
for the parton pairs $\{i,j\}$ and are defined as:
\begin{equation} W_{ij}=\frac{p_i\cdot p_j}{(p_i\cdot p_g)(p_j\cdot p_g)}
\end{equation}
for the emission of a soft gluon of momentum $p_g$. In terms of the angles
between parton pairs,
\begin{equation}W_{ij}=\frac{1}{E_g^2}\cdot
\frac{\xi_{ij}}{\xi_{ig}}{\xi_{jg}}\end{equation}
where $\xi_{ij}=1-\cos\theta_{ij}=p_i\cdot p_j/E_iE_j$. $E_g$ is the soft
gluon energy. This expression contains poles at $\theta_{ig}=0$ and
$\theta_{jg}$. These correspond to the two collinear directions in gluon
emission. Explicitly,
\begin{equation}W_{ij}=W^i_{ij}+W^j_{ij}=
\frac{1}{2E_g^2}\left\{\frac{\xi_{ij}}{\xi_{ig}\xi_{jg}}+
\frac{1}{\xi_{ig}}-\frac{1}{\xi_{jg}}\right\}+\{i\lr j\}.
\end{equation}
$W^i_{ij}$ have collinear singularities only at $\theta_{ig}=0$. It can be
shown that the azimuthal average of $W^i_{ij}$ around the $i$-th parton
direction is a step function with cut-off at $\theta_{ij}$, hence the
above claim concerning the radiation pattern.

We now consider modifying the planar terms such that the sum of them is
equal to the original matrix element squared. Note that:
\begin{eqnarray}
W_t-W_\np &=& \ca \left[W_{12}+W_{34}-W_{14}-W_{23}\right]\\
W_u-W_\np &=& \ca \left[W_{12}+W_{34}-W_{13}-W_{24}\right].
\end{eqnarray}
In both expressions the radiation cancels in all four collinear
directions, and so it is reasonable to approximate $W_\np$ by either $W_t$
or $W_u$.

Hence it can be deduced that the radiation due to the $(1/\nc^2)$
suppressed \np\ term can be treated approximately by distributing
this term between the two colour flows in some ratio.

From the viewpoint of practicality in \MC\ simulations, this
distribution should be such that the modified planar terms, which are
called `full terms' in \cite{MW}, should be positive definite.

\section{The MW procedure}\label{sec_MW}

The procedure of Marchesini and Webber (MW) for evaluating the full terms
is as follows \cite{MW}:
\begin{itemize}
\item the full term should have the same pole structure and crossing
symmetry as the planar term;
\item it should remain positive definite in order to be interpreted as a
probability distribution;
\item the sum of the full terms should give the exact lowest order $2\to2$
matrix element squared.
\end{itemize}

The second and the third criteria above are essential as argued earlier. 
As for the pole structure, the introduction of an extra physical pole is
bound to drive at least one of the full terms negative for some
permutation of the external legs so this is a natural consequence of the
second criterion. The requirement of correct crossing symmetry is never
utilised in practice, and it is in fact too constraining to be
practicable, as will be shown later. Discarding this requirement, we have
essentially only the positivity and the sum to constrain the colour
rearrangement, which is not sufficient to determine it uniquely. One might
therefore introduce the following criterion, which is similar to the
requirement of (not only physical) pole structure, and can be regarded as
being inherent in the MW procedure: 
\begin{itemize}
\item when a \np\ term contains poles corresponding to two colour
flows, this is split up by partial fractions, viz
\begin{equation}\frac{1}{st}+\frac{1}{su}+\frac{1}{tu}=0\end{equation}
for massless partons.
\end{itemize}

There are several disadvantages associated with this method:
\begin{itemize}
\item the decomposition of the \np\ part is not unique without the
additional criterion introduced above, and with this criterion the
procedure sometimes fails and/or is still not unique, depending on the
exact procedure by which the partial fractions are split up; 
\item the decomposition is not general under permutations of external 
legs;
\item the decomposition becomes laborious when the number of colour flows
is increased, and verification becomes an impossible task. The number
of colour flows is a factorial function of the number of external partons.
\end{itemize}
The first two of these points are illustrated well for the case of the
\susy\ \qcd\ process $gg\to\squark\squark^*$ which, at the tree
level and in the massless limit, is described by the following matrix
element squared:
\begin{equation} \M = \frac{\gsf\nc}{\nc^2-1}\cdot
\frac{(u^2)_t+(t^2)_u+(-s^2/\nc^2)_\np}{s^2}. \end{equation}
This does not have a unique rearrangement, not all such decompositions
are positive, and many of them are not invariant under permutations of
the external legs.

As a further illustration of this second point, consider the process
$qq\to qq$, which has two colour flow structures as shown in figure
\ref{figB}, and whose matrix element squared is rearranged, according to
\cite{MW}, as follows:
\begin{eqnarray} \M &=& \frac{\gsf\cf}{\nc}
\left[\left(\frac{s^2+u^2}{t^2}\right)+\left(\frac{s^2+t^2}{u^2}\right)+
\left(-\frac{2}{\nc}\cdot\frac{s^2}{ut}\right)\right]\\\label{MWprocB}
&=& \frac{\gsf\cf}{\nc}
\left[\left(\frac{s^2+u^2}{t^2}+\frac{2}{\nc}\cdot\frac{s}{t}\right)
+\left(u\lr t\right) \right]. \end{eqnarray}

This arrangement does not preserve the crossing symmetries $(s\lr u)$ and
$(s\lr t)$ of the planar terms. It is in fact impossible to preserve the
crossing symmetry while also preserving the pole structure, since the only
dimensionless combination of $s$, $t$ and $u$ that is $(s\lr u)$ symmetric
is $su/t^2$. Taking into account the pole structure, the \np\ term
$s^2/ut$ can not be expressed as a linear combination of $su/t^2$,
$st/u^2$ and a constant.

Although (\ref{MWprocB}) seems a natural rearrangement, it fails when
$\nc=2$ and the leg permutation $(s\lr u)$ is made,
corresponding to $q\bar{q}\to q\bar{q}$: 
\begin{equation} \M = \frac{\gsf\cf}{\nc}\left[
\left(\frac{s^2+u^2}{t^2}+\frac{2}{\nc}\cdot\frac{u}{t}\right)+
\left(\frac{u^2+t^2}{s^2}+\frac{2}{\nc}\cdot\frac{u}{s}\right)
\right].\end{equation}
If we now set $\nc=2$ in the second term above, we obtain
\begin{equation}
\frac{u^2+t^2}{s^2}+\frac{u}{s}=\frac{(-t)(u-t)}{s^2}\end{equation}
which is negative whenever $(u-t)$ is negative. For $\nc=3$ the above
expression remains positive but this is accidental\footnote{In general,
when a full term is positive for a certain \nc, one can prove trivially
that it is also positive for all higher \nc\ since the planar term must
always be positive.}.
\begin{equation}
u^2+t^2+\frac{2}{3}su = \frac{1}{3}(u-t)^2+\frac{2}{3}t^2 > 0.
\end{equation}

One method for obtaining a positive definite decomposition is to split the
\np\ term as follows:
\begin{equation}
-\frac{u^2}{st} = \frac{u^2}{s-t}\left(-\frac{1}{t}+\frac{1}{s}\right).
\end{equation}
Since both contributions are positive, the full terms are also positive. 
However, the fact remains that the decomposition is not universal under
permutations of the external legs.

\section{Proposed new procedure}\label{sec_new}

To solve the above problems, we propose the following procedure. Let the
overall matrix element squared be given by:
\begin{equation}
\M_{tot} = \left(\sum_i \M_i\right) + \left(\np\right)
       = \M_{\rm planar} + \left(\np\right). \end{equation}
where $\M_i$ is the planar term for the $i$-th colour flow, and $(\np)$
represents the \np\ part. Then the $i$-th full term is given by:
\begin{eqnarray}\M_{{\rm full},i}
&=& \frac{\M_i}{\M_{\rm planar}} \M_{\rm tot} \label{newdecomp1}\\
&=& \M_i+\frac{\M_i}{\M_{\rm planar}} \left(\np\right).
\label{newdecomp2} \end{eqnarray}
In other words, we split the \np\ part by the ratios of the planar
terms.

This is positive definite, as is obvious in equation (\ref{newdecomp1}). 
This method resolves all three drawbacks of the MW method listed above,
and carries some additional advantages:
\begin{itemize}
\item the correct pole structure in each full term is automatically
ensured --- the larger the planar term the larger the full term. No extra
pole is introduced since the sum of the planar terms is positive definite
(even in unphysical regions) and furthermore can not approach zero faster
than the \np\ part; 
\item related processes always have the same decomposition;
\item we do not even need a compact analytical expression for the matrix
elements squared to carry out the colour decompositions.
\end{itemize}
This last feature becomes useful for more complex $n$-body
processes\footnote{In \cite{LEP2} a variant of this procedure is adopted
as the default in the process $e^+e^-\to q\bar{q}gg$.}, where the most
efficient method for calculating matrix elements may be to utilise the
helicity amplitude formalisms \cite{HZ,HELAS}.

One drawback of this method is that the formulae may not always be
aesthetically preferable to those obtained with the method of Marchesini
and Webber, for example in the case of\footnote{In fact, for all other
massless \qcd\ processes the procedure described here yields results
identical to those in \cite{MW}.} $qq\to qq$. However, this is not a
problem in computer simulations, which are the only circumstances where
the methods described in this paper are put to practice. 

\section{The 2 parton $\to$ 2 sparton processes}\label{sec_susy}

The gauge invariance and the simplicity of these processes greatly
compactify the expressions for matrix elements. In particular, as in the
\qcd\ case, all \susy\ processes of the form $q\bar{q}\to gg$ (where $q$
and $g$ refer here to quarks or squarks and gluons or gluinos,
respectively) can be expressed as: 
\begin{equation}{\rm (colour\ factor)}\times
\frac{(u_4^2)_t+(t_3^2)_u+(-s^2/\nc^2)_\np}{s^2}\times
\M_{\rm QED} \end{equation}
provided that the sparticles exchanged in the $t$ and the $u$ channels
have masses $m_{3(4)}$ and $m_{4(3)}$ respectively for the process $1+2\to
3+4$. Here $u_4=u-m_4^2=-2p_1\cdot p_4$ and $t_3=t-m_3^2=-2p_1\cdot p_3$.
When $m_t\neq m_{3(4)}$ or $m_u\neq m_{4(3)}$, as is the case generally
in $q\bar{q}\to\gluino\gluino$, there are correction terms proportional to
the differences in the squared masses.

The colour flows are identical to the ordinary \qcd\ case shown in figures
\ref{figA}--\ref{figD}.

We present the spin and colour averaged squared matrix elements, general
for any \nc\ and nondegenerate squark masses. These are divided by a
statistical factor of two when the final state spartons are identical.

Apart from the colour structures, the formulae match those of \cite{DEQ}
when $N_C=3$, and those of tree-level expressions in \cite{BHSZ} when left
and right squark masses are taken to be degenerate. Stop and sbottom
mixings are not considered here since we are dealing with chirality
independent interactions.

In the formulae that follow, $g_s$ is the strong coupling, evaluated at
some scale which is not determined at tree level (in the \herwig\ \MC\ it
is taken to be $\sqrt{s}$). $s$ is the effective centre-of-mass energy
$s_{\rm eff} = s_{\rm tot} x_1 x_2$, $t$ and $u$ similarly. Since we are
taking the initial state partons to be massless it follows that
$s+t+u=m_3^2+m_4^2$ and $s+t_3+u_4=0$. Also $ut-m_3^2m_4^2 = sp_T^2\geq
0$, where $p_T$ is the outgoing transverse momentum. $\mgluino$ is the
gluino mass, $\MiLR$ are the $i$-th generation squark masses. Charge
conjugate final states must also be included in simulations. 
\begin{equation}
\M(q_i\bar{q}_i\to\gluino\gluino)=\frac{g^4_sC_F}{4}\sum_{L,R}C_{L,R}
\end{equation}
with
\begin{multline} C_{L,R}=\frac{2sp_T^2}{s^2}\left[
\frac{\left(u_4^2-\Delta^2\right)_t+\left(t_3^2-\Delta^2\right)_u-
\left(s^2/\nc^2\right)_\np}{(u_4-\Delta)(t_3-\Delta)}
\right]\\
+\Delta^2\left[\left(\frac{1}{(t_3-\Delta)^2}\right)_t+
\left(\frac{1}{(u_4-\Delta)^2}\right)_u-\frac{1}{\nc^2}\left(
\frac{1}{t_3-\Delta}-\frac{1}{u_4-\Delta}\right)^2_\np\right]
\end{multline}
where $\Delta = \MiLR^2-\mgluino^2$. $\sum_{L,R}$ denotes a summation over
the left and right squarks. Terms marked with subscript $t$ correspond to
the colour flow $(1\to3,3\to4,4\to2)$ and those with subscript $u$
correspond to the colour flow $(1\to4,4\to3,3\to2)$, as in figure
\ref{figC}. Although it is possible to distribute the \np\ terms between
the two colour flows using the `intuitive' method, for the sake of
consistency we advocate the use of equation (\ref{newdecomp2}). 

\begin{multline}
\M(gg\to\gluino\gluino)
=\frac{g^4_sN_C^2}{N_C^2-1}\cdot\frac{u_4t_3}{2}
\left[ u_4^2+t_3^2+\frac{4\mgluino^2s^2p_T^2}{u_4t_3} \right]\\\times
\left[\left(\frac{1}{s^2t_3^2}\right)_{st}+
\left(\frac{1}{s^2u_4^2}\right)_{su}+
\left(\frac{1}{u_4^2t_3^2}\right)_{ut}\right]
\end{multline}
The colour flows are $(1\to3,3\to4,4\to2,2\to1)$ for $1/s^2t_3^2$,
$(1\to4,4\to3,3\to2,2\to1)$ for $1/s^2u_4^2$, and
$(1\to4,4\to2,2\to3,3\to1)$ for $1/t_3^2u_4^2$ (figure \ref{figD}). There
is no \np\ term, as any cross term between two colour flows is equivalent
to the square of the other. 

\begin{multline}
\M(gq_i\to\gluino\squark_{i_{L,R}}) = \frac{g^4_s}{4}
\left[ -u_4-2(m_4^2-m_3^2)\left(1+\frac{m_3^2}{t_3}+\frac{m_4^2}{u_4}
\right)\right]\\
\times\frac{(u_4^2)_s+(s^2)_u-(t_3^2/\nc^2)_\np}{st_3u_4}
\end{multline}
The colour flows are $(2\to1,1\to3,3\to4)$ for the $s$-channel term and
$(2\to3,3\to1,1\to4)$ for the $u$-channel term (figure \ref{figC}). 
Equation (\ref{newdecomp2}) leads to the decomposition of the \np\ term
$t_3^2 = [(u_4^2t_3^2)_s+(s^2t_3^2)_u]/(u_4^2+s^2)$. 

\begin{multline}
\M(q_iq_j\to\squark_{i_{L,R}}\squark_{j_{L,R}})=
\frac{g^4_sC_F}{2N_C}\cdot\frac{\mgluino^2s}{1+\delta_{ij}}\\\times
\left[\left(           \frac{1}{(t-\mgluino^2)^2}\right)_t+
      \left(\delta_{ij}\frac{1}{(u-\mgluino^2)^2}\right)_u-
      \left(\delta_{ij}
       \frac{2/\nc}{(t-\mgluino^2)(u-\mgluino^2)}\right)_\np
\right]
\end{multline}
The colour flows are $(1\to4,2\to3)$ for the $t$-channel term and
$(1\to3,2\to4)$ for the $u$-channel term (figure \ref{figB}). For the case
$i\neq j$, the colour flow is uniquely $(1\to4,2\to3)$, as in figure
\ref{figA}.

\begin{multline}
\M(q_iq_j\to\squark_{i_{L,R}}\squark_{j_{R,L}})=\frac{g^4_sC_F}{2N_C}
\cdot sp_T^2\\\times\left[
\left(\frac{1}{(t-\mgluino^2)^2}\right)_t+
\left(\delta_{ij}\frac{1}{(u-\mgluino^2)^2}\right)_u\right]
\end{multline}
The colour flows are again $(1\to4,2\to3)$ for the $t$-channel term and
$(1\to3,2\to4)$ for the $u$-channel term. There is no \np\ term. For
the case $i\neq j$, the colour flow is again uniquely $(1\to4,2\to3)$. 

\begin{multline}
\M(q_i\bar{q}_j\to\squark_{i_{L,R}}\squark_{j_{L,R}}^*)
=\frac{g^4_sC_F}{2N_C}\cdot sp_T^2\\\times\left[
   \left(\frac{1}{(t-\mgluino^2)^2}\right)_t+
   \left(\delta_{ij}\frac{2}{s^2}\right)_s-
   \left(\delta_{ij}\frac{2/\nc}{s(t-\mgluino^2)}\right)_\np
 \right]
\end{multline}
The colour flows are $(1\to2,4\to3)$ for the $t$-channel term and
$(1\to3,4\to2)$ for the $s$-channel term. For the case $i\neq j$, the
colour flow is uniquely $(1\to2,4\to3)$. 

\begin{equation}
\M(q_i\bar{q}_j\to\squark_{i_{L,R}}\squark_{j_{R,L}}^*)=\frac{g^4_sC_F}{2N_C}
\cdot\frac{\mgluino^2s}{(t-\mgluino^2)^2}
\end{equation}
The colour flow is uniquely $(1\to2,4\to3)$, regardless of $i$ and $j$.

\begin{equation}
\M(q_i\bar{q}_i\to\squark_{j_{L,R}}\squark_{j_{L,R}}^*)=\frac{g^4_sC_F}{N_C}
\cdot\frac{sp_T^2}{s^2}
\end{equation}
Here $i\neq j$. The colour flow is uniquely $(1\to3,4\to2)$.

\begin{multline}
\M(gg\to\squark_{i_{L,R}}\squark_{i_{L,R}}^*)
=\frac{g^4_sN_C}{2(N_C^2-1)}\left[(sp_T^2)^2+m_3^2m_4^2s^2\right]\\ \times
\frac{(u_4^2)_t+(t_3^2)_u-(s^2/\nc^2)_\np}{s^2t_3^2u_4^2}
\end{multline}
The colour flows are $(4\to2,2\to1,1\to3)$ for the $t$-channel term and
$(4\to1,1\to2,2\to3)$ for the $u$-channel term (figure \ref{figC}).

\section{Conclusions}\label{sec_concl}

We have discussed the techniques for calculating the `full terms' in \qcd\
and \susy\ \qcd\ processes which involve multiple colour flows. The exact
distribution of \np\ terms between the full terms is expected to make
little significant difference to the prediction of soft radiation
patterns, but we argued that the conventional method used to achieve this
is inadequate.

We have presented the formulae for 2 parton $\to$ 2 sparton processes,
together with the colour flows associated with them. These are
incorporated in the \MC\ event generator \herwig\ 6.1, to be released
shortly. However, the study of the showering and hadronisation of \susy\
particles has not been carried out in sufficient detail, and it is clear
that further investigations are necessary.

\section*{Acknowledgements}

I thank Bryan Webber for suggesting the study of colour decompositions of
\susy\ processes to me and for the fruitful discussions, Stefano Moretti
for continuous encouragement and for testing some of my earlier attempts
at colour decompositions, and Pino Marchesini for helpful comments. I
acknowledge the financial support from Trinity College and the Committee
of Vice-Chancellors and Principals of the Universities of the UK.

\newpage\section*{Figure captions}
\renewcommand{\labelenumi}{[\theenumi]}\begin{enumerate}
\item\label{figA} Colour flow for processes of the type
$q_1q'_2\to q_3q'_4$.
\item\label{figB} Colour flows for processes of the type
$q_1q_2\to q_3q_4$.
\item\label{figC} Colour flows for processes of the type
$q_1\bar{q}_2\to g_3g_4$.
\item\label{figD} Colour flows for processes of the type
$g_1g_2\to g_3g_4$. The direction of colour flows (indicated by the
arrows) is arbitrary.
\end{enumerate}

\newpage \thispagestyle{empty} \setlength{\unitlength}{4.3mm}
   \newcommand{\colA}[5]{ \begin{picture}(10,10) \put(4.5,0.5){`$#1$'}
\put(0.8,8){$#2$}\put(0.8,2){$#3$}\put(8.6,8){$#4$}\put(8.6,2){$#5$}
\put(2.5,8){\vector(1,0){1}} \put(2,8){\line(1,0){2}}
\put(2.5,2){\vector(1,0){1}} \put(2,2){\line(1,0){2}}
\put(6  ,8){\vector(1,0){1}} \put(6,8){\line(1,0){2}}
\put(6  ,2){\vector(1,0){1}} \put(6,2){\line(1,0){2}}
\put(4  ,8){\line( 1,-3){2}} \put(4,2){\line(1,3){2}} \end{picture} }
   \newcommand{\colC}[3]{ \begin{picture}(10,10) \put(4.5,0.5){`$#1$'}
\put(1.1,8){$q_1$}\put(1.1,2){$q_2$}\put(8.9,8){$g_#2$}\put(8.9,2){$g_#3$}
\put(2.5,8){\vector( 1, 0){1}} \put(2.5,8){\vector( 1,0){5}}
\put(8  ,2){\vector(-1, 0){1}} \put(8  ,2){\vector(-1,0){5}}
\put(2  ,8){  \line( 1, 0){6}} \put(8  ,2){  \line(-1,0){6}}
\put(8  ,7){\vector(-1, 0){2}} \put(8  ,7){  \line(-1,0){3}}
\put(5.5,3){\vector( 1, 0){1}} \put(5  ,3){  \line( 1,0){3}}
\put(5  ,7){  \line( 0,-1){4}} \end{picture} }
   \newcommand{\colD}[4]{ \begin{picture}(10,10) \put(4.3,0.5){`$#1$'}
\put(1.4,8){$g_1$}\put(1.4,2){$g_#2$}\put(7.8,8){$g_#3$}\put(7.8,2){$g_#4$}
\put(3,8){\vector( 1,-1){1}} \put(4,7){\line( 1,-1){1}}
\put(4,5){\vector(-1, 1){1}} \put(3,6){\line(-1, 1){1}}
\put(2,3){\vector( 1, 1){1}} \put(3,4){\line( 1, 1){1}}
\put(5,4){\vector(-1,-1){1}} \put(4,3){\line(-1,-1){1}}
\put(5,6){\vector( 1, 1){1}} \put(6,7){\line( 1, 1){1}}
\put(8,7){\vector(-1,-1){1}} \put(7,6){\line(-1,-1){1}}
\put(6,5){\vector( 1,-1){1}} \put(7,4){\line( 1,-1){1}}
\put(7,2){\vector(-1, 1){1}} \put(6,3){\line(-1, 1){1}} \end{picture} }
\begin{center} \colA{t}{q_1}{q'_2}{q_3}{q'_4}\\Figure \ref{figA}\\
\colA{t}{q_1}{q_2}{q_3}{q_4}\colA{u}{q_1}{q_2}{q_4}{q_3}\\Figure \ref{figB}\\
\colC{t}{3}{4}\colC{u}{4}{3}\\Figure \ref{figC}\\
\colD{st}{2}{3}{4}\colD{su}{2}{4}{3}\colD{ut}{4}{3}{2}\\Figure \ref{figD}
\end{center}
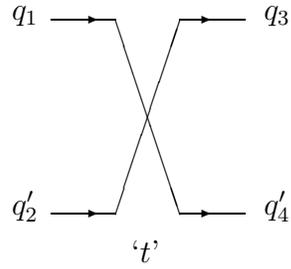
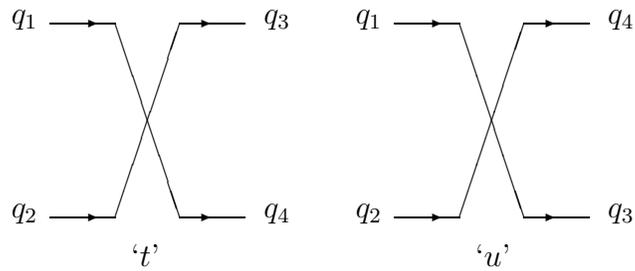
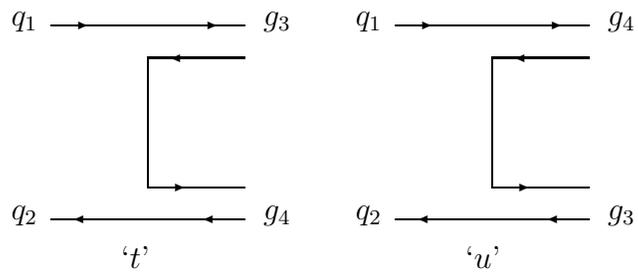
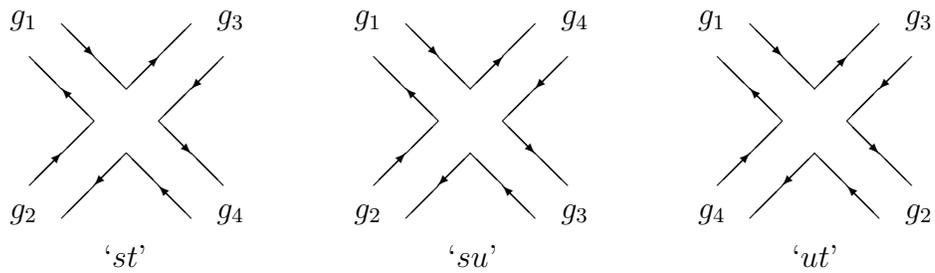
\end{document}